\documentclass[12pt]{article}
\usepackage{latexsym}
\usepackage[dvips]{graphics}
\begin{document}

\title{Theorems on gravitational time delay and related issues}

\author{Sijie Gao and  Robert M. Wald \\Enrico Fermi Institute and
Department of Physics \\   
University of Chicago \\
5640 S. Ellis Avenue \\
Chicago, Illinois 60637-1433}

\maketitle

\begin{abstract}
Two theorems related to gravitational time delay are proven.  Both
theorems apply to spacetimes satisfying the null energy condition and
the null generic condition. The first theorem states that if the
spacetime is null geodesically complete, then given any compact set
$K$, there exists another compact set $K'$ such that for any $p,q
\not\in K'$, if there exists a ``fastest null geodesic'', $\gamma$,
between $p$ and $q$, then $\gamma$ cannot enter $K$. As an application
of this theorem, we show that if, in addition, the spacetime is
globally hyperbolic with a compact Cauchy surface, then any observer
at sufficiently late times cannot have a particle horizon. The
second theorem states that if a timelike conformal boundary can be
attached to the spacetime such that the spacetime with boundary
satisfies strong causality as well as a compactness condition, then
any ``fastest null geodesic'' connecting two points on the boundary
must lie entirely within the boundary. It follows from this theorem
that generic perturbations of anti-de Sitter spacetime always produce a
time delay relative to anti-de Sitter spacetime itself.

\end{abstract}

\section{Introduction}

Gravitational time delay effects in general relativity have been
considered by many authors. Although general relativity makes
unambiguous predictions for, e.g., the time of reception of signals
sent by a space probe orbiting nearly behind the sun---and these
predictions have been observationally verified \cite{r}---it appears
remarkably difficult to characterize these effects in a meaningful way
in terms of the ``time delay'' of a light ray relative to the time
required in Minkowski spacetime. The reason for this difficulty is
that, in general, there is no natural way to choose a flat background
for a curved spacetime, and, thus, no meaningful way to compare the
propagation of a light ray in a curved spacetime with that of a
``corresponding'' light ray in Minkowski spacetime \cite{w}.

Recently, a general result on gravitational time delay was obtained in
\cite{vbl}. The authors of \cite{vbl} considered linearized
perturbations
\begin{equation}
h_{ab}= g_{ab} - \eta_{ab}
\end{equation}
of Minkowski spacetime that satisfy the linearized version of the
the null energy condition. (The null energy condition states
that
\begin{equation}
R_{ab}k^ak^b \geq 0
\label{nec}
\end{equation}
for all null $k^a$.) It was further required that no incoming
gravitational radiation be present. It was then shown that in the Lorentz
gauge,
\begin{equation}
\partial_b [h^{ab}-\frac{1}{2}\eta^{ab} h]=0
\end{equation}
the metric perturbation satisfies
\begin{equation}
h_{ab}k^a k^b \geq 0
\label{hk}
\end{equation}
for all $k^a$ that are null with respect to the Minkowski background,
i.e., $\eta_{ab} k^a k^b = 0$. Eq.(\ref{hk}) can be interpreted as saying
that the light cones of the perturbed metric contract relative to the
flat background, i.e., the propagation of light is ``slower'' in the
perturbed spacetime.

However, the above result is highly gauge dependent, i.e., it depends
in a crucial way on how one chooses to compare the perturbed spacetime
with the background Minkowski spacetime. The choice of a different
manner of comparison corresponds to changing $h_{ab}$ via a gauge
transformation
\begin{equation}
h_{ab}\rightarrow h_{ab}+\partial_a \xi_b+\partial_b \xi_a
\end{equation}
The following is an explicit example of a smooth, pure gauge $h_{ab}$
that ``opens out'' the light cones everywhere and also goes to zero as
$1/r$ at spatial and null infinity: In spherical polar coordinates in
Minkowski spacetime, let
\begin{equation}
\xi_a= -\frac{r_a}{2+g(r)}
\end{equation}
where $r^a$ is the radially outward pointing vector with $r^a r_a =
r^2$ (i.e., $r_a = r (dr)_a$) and $g$ is any smooth function such that
$g(r) = r$ for $r \geq 1$, $g(r) = r^2$ for $0 \leq r \leq 1/2$, and
$g(r) \geq 0$, $0 \leq g^\prime (r) < 2$ for all $r$. Then, we have
\begin{eqnarray}
h_{ab} &=& 2 \partial_{(a} \xi_{b)} \nonumber \\
&=&  \frac{2g'(r)}{[2+g(r)]^2} (dr)_a r_b 
- \frac{2}{2+g(r)} \partial_{(a} r_{b)} \nonumber \\
&=& \frac{2r g'(r)}{[2 + g(r)]^2} (dr)_a (dr)_b - \frac{2}{2+g(r)} q_{ab}
\end{eqnarray}
where $q_{ab}$ is the Euclidean metric on the $t = \rm{constant}$
hypersurfaces. Since $g(r) = r$ for $r \geq 1$, we see that the
orthonormal frame components of $h_{ab}$ are $O(1/r)$ as $r \rightarrow
\infty$ (independently of $t$). For any null vector $k^a$, we have
\begin{eqnarray}
h_{ab} k^a k^b &=& \frac{2r g'(r)}{[2 + g(r)]^2} (\frac{\vec k\cdot \vec x}{r})^2 
- \frac{2}{2+g(r)} |\vec k|^2 \nonumber \\
&<& 0
\end{eqnarray}
i.e., this pure gauge $h_{ab}$ opens out the light cones
everywhere. By adding a suitable multiple of this pure gauge
perturbation to the perturbation considered in \cite{vbl}, we can
reverse the conclusions of \cite{vbl} in any compact region of
spacetime.

Indeed, a theorem of Penrose \cite{p} proves that it is impossible to
identify Schwarzschild spacetime with Minkowski spacetime so that the
null infinities of the two spacetimes coincide and the light cones of
the Schwarzschild metric lie within the lightcones of the Minkowski
metric. (This does not contradict the results of \cite{vbl} because
the Lorentz gauge is ill behaved at null infinity \cite{gx}.) From the
point of view taken in \cite{vbl}, Penrose's result might be
interpreted as an ``anti-time-delay'' theorem, but it would seem more
reasonable to interpret both Penrose's theorem and the results of
\cite{vbl} as having more to do with the manner in which the
identification of the curved and flat spacetimes are made than with
gravitational time delay.

As pointed out by Olum \cite{o}, in the case of a spacetime that is
Minkowskian outside of a world tube (see Fig. 2 of \cite{o}), an
unambiguous comparison with Minkowski spacetime can be done in the
Minkowskian region, so the notion of ``time delay'' is well defined in
this context. It was shown in \cite{o} that in such a spacetime, if
the null energy condition (\ref{nec}) and null generic condition (see
eq.(\ref{gc}) below) hold within the non-flat region, then a ``time
advance'' cannot occur. However, if the dominant energy condition
holds, it is impossible to have a spacetime (other than Minkowski
spacetime) that is Minkowskian outside of a worldtube, since the ADM
mass of such a spacetime would vanish, in contradiction with the
positive mass theorem. In \cite{o}, an alternative characterization of
``time delay'' also was proposed. This characterization does not
require a comparison with Minkowski spacetime, and a time delay
theorem was proven for this definition.

In this paper, we will prove two theorems related to ``time delay''
that also do not require a comparison with Minkowski spacetime. Both
theorems apply to spacetimes that satisfy the null energy condition
(\ref{nec}) as well as the null generic condition. (The null generic
condition is the statement that each null geodesic contains a point at
which
\begin{equation}
k_{[a} R_{b]cd[e} k_{f]} k^c k^d \neq 0
\label{gc}
\end{equation}
where $k^a$ denotes the tangent to the geodesic.) It appears likely
that our assumption that the null generic condition holds could be
significantly weakened by use of the null splitting theorems of
\cite{g}, but we shall not consider this issue further here.

Our first theorem states that in a null geodesically complete
spacetime satisfying the null energy condition and the null generic
condition, given any compact set $K$, there exists another compact set
$K'$ satisfying the following property: If $p$ and $q$ are any two
events lying outside of $K'$ such that $q \in J^{+}(p)-I^{+}(p)$, then
any causal curve, $\gamma$, connecting $p$ with $q$ (which necessarily
must be a null geodesic) cannot enter $K$ (see Fig. 1). The
relationship of this result to ``time delay'' is as follows: If it
were possible to arrange the spacetime geometry in a compact region
$K$ so as to produce a ``time advance'', then one might expect a
``fastest null geodesic'' $\gamma$ to take advantage of this by
entering $K$. Thus, the theorem suggests that ``time advance'' is not
possible, although it is difficult to make a strong argument for this
interpretation, since the theorem gives little control over the size
of the region $K'$. Some additional applications of this
theorem will be given in section 2.

\begin{figure}[t]
\resizebox{\textwidth}{!}
{\includegraphics[-2.5in,0in][9.5in,6in]{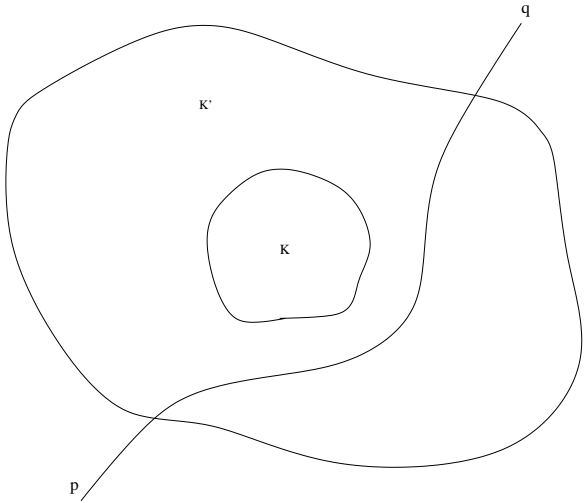}}
\caption{In a null geodesically complete spacetime satisfying the null
energy condition and the null generic condition, given a compact
region $K$, there exists another compact region $K^\prime$ such that
if $p,q \not\in K'$ and $q \in J^+(p)-I^+(p)$, then any causal
curve connecting $p$ to $q$ cannot intersect $K$.}
\end{figure}

In section 3, we present our second theorem. In this theorem, we
restrict attention to spacetimes satisfying the null energy condition
and null generic condition to which a timelike conformal boundary can
be attached, such that the conformally completed spacetime, $\bar{M}$,
satisfies strong causality together with the property that for all
$p,q \in \bar{M}$, the set $J^+(p) \cap J^-(q) \cap \bar{M}$ is
compact. The prototype spacetimes satisfying these properties are
asymptotically anti-de Sitter spacetimes, although anti-de Sitter
spacetime itself would not be in this class on account of its failure
to satisfy the null generic condition. For the spacetimes satisfying
the hypotheses of this theorem, we prove that any ``fastest null
geodesic'' in $\bar{M}$ connecting two points $p,q$ in the conformal
boundary, $\dot{M}$, must lie entirely within $\dot{M}$. Now, in the
case of anti-de Sitter spacetime itself, it turns out that if the
points $p,q \in \dot{M}$ are the past and future endpoints in
$\bar{M}$ of a null geodesic in $M$, then there also exists a null
geodesic in $\dot{M}$ connecting $p$ and $q$, but no ``faster'' null
geodesic exists, i.e., $q$ lies on the boundary of the future of $p$
in $\bar{M}$. In other words, in anti-de Sitter spacetime, in a
``race'' to get from $p \in \dot{M}$ to a point in $\dot{M}$ antipodal
to $p$, there is a ``tie'' between null geodesics passing through $M$
and null geodesics lying in $\dot{M}$. Our theorem proves that in a
generically perturbed anti-de Sitter spacetime, this ``tie'' is always
broken in favor of geodesics lying within the boundary. Thus, our
theorem can be interpreted as saying that in generically perturbed
anti-de Sitter spacetime, there is always a ``time delay'' relative to
anti-de Sitter spacetime itself. A similar result under somewhat
different hypotheses has previously been obtained by Woolgar
\cite{wo}.

Our notation and conventions throughout this paper follow those of
\cite{w}. All spacetimes considered in this paper will be assumed to
be connected, smooth, time oriented, and paracompact. No assumption
will be made about the dimensionality of spacetime, i.e., our results
hold in any spacetime dimension.

\section{Avoidance of Compact Sets by Sufficiently Long ``Fastest Null 
Geodesics''}

In this section, we will prove the following theorem\footnote{Recall
that a {\em null line} is an inextendible, achronal null
geodesic. G. Galloway (private communication) has pointed out to us
that the conclusions of Theorem 1 remain valid if the hypothesis of
Theorem 1 is replaced by the condition that $(M, g_{ab})$ fails to
contain a null line. Since the present hypotheses of Theorem 1 (i.e.,
null geodesic completeness, the null energy condition, and the null
generic condition) preclude the existence of a null line, Galloway's
modification of Theorem 1 is a stronger result than our
version. Galloway's proof of his version of Theorem 1 provides a
direct construction of a null line when the conclusion of Theorem 1
fails, thereby bypassing Lemma 1 (whose proof makes explicit use of
the null energy condition).\label{gg}} and then discuss some
applications of it.

\newtheorem{thm}{Theorem}
\begin{thm}
Let $(M, g_{ab})$ be a null geodesically complete spacetime
satisfying the null energy condition (\ref{nec}) and the null generic
condition (see eq.(\ref{gc})). Then given any compact region $K
\subset M $, there exists another compact region $K^\prime$ containing
K such that if $q,p \notin K^\prime$ and $q \in J^{+}(p)-I^{+}(p)$,
then any causal curve $\gamma$\ connecting $p$ to $q$ cannot intersect
the region $K$.
\end{thm}

The proof of this theorem, will be based upon a lemma that we shall
state and prove below. Before doing so we introduce some notation and
recall some facts about conjugate points. Since $M$ is time oriented,
we may choose a continuous, nowhere vanishing, future-directed
timelike vector field $t^a$. We define
\begin{equation}
{\cal S} = \{(p,k^a)|  p\in M, k^a \in V_p, k^a k_a = 0, k^a t_a = -1 \}
\label{S}
\end{equation}
where $V_p$ denotes the tangent space at $p$.  Thus, $\cal S$ consists
of the points in the tangent bundle of $M$ where the tangent vector is
a future directed null vector that is normalized with respect to
$t^a$. We shall denote points of $\cal S$ by $\Lambda$. Associated
with any $\Lambda = (p, k^a) \in {\cal S}$ is the null geodesic
$\gamma_\Lambda$ which starts at $p$ with tangent $k^a$; we take the
affine parameter, $\lambda$, of $p$ along $\gamma_\Lambda$ to be $0$,
i.e., $\gamma_\Lambda (0) = p$.

Recall that points $p,q$ along a geodesic $\gamma$ are said to be
conjugate if there exists a Jacobi field which vanishes at both $p$
and $q$. Equivalently (see, e.g., \cite{he} or \cite{w}), if we define
the matrix ${A^{\mu}}_{\nu}(\lambda)$ at affine parameter $\lambda$
along $\gamma$ by
\begin{equation}
\frac{d^2 {A^{\mu}}_{\nu}}{d^2 \lambda}=-\sum_{\alpha,\beta,\sigma} {R_{\alpha\beta\sigma}}^{\mu}k^{\alpha} k^{\sigma}{A^{\beta}}_{\nu}
\label{A}
\end{equation}
with initial conditions ${A^{\mu}}_{\nu}|_p = 0$ and
$(d{A^{\mu}}_{\nu}/d\lambda) |_p={\delta^{\mu}}_{\nu}$, then $q$ will
be conjugate to $p$ if and only if $\det A = 0$ at $q$. We define
\begin{equation}
G(\lambda) = \sqrt{\det A(\lambda)}
\label{Gdef}
\end{equation}
Since ${A^{\mu}}_{\nu}$ is a solution to the linear ordinary
differential equation (\ref{A}), it follows that ${A^{\mu}}_{\nu}$ and
hence $\det A$ vary smoothly with $(\Lambda, \lambda)$. Consequently,
$G$ varies smoothly with $(\Lambda, \lambda)$ except possibly at
points where $G=0$. Finally, we recall that in terms of $G$, the
Raychaudhuri equation yields (see, e.g., \cite{fmw})
\begin{equation}
G''/G = - \frac{1}{2} [\sigma_{ab} \sigma^{ab} + R_{ab} k^a k^b]
\label{ray}
\end{equation}
where $\sigma_{ab}$ denotes the shear of the congruence of null
geodesics emanating from $p$.

\newtheorem{lem}{Lemma}
\begin{lem}  Let $(M, g_{ab})$ satisfy the null energy condition 
(but not necessarily be null geodesically complete nor satisfy the
null generic condition). Let $\Lambda_0 = (p_0, k_0^a) \in {\cal S}$
be such that the null geodesic $\gamma_{\Lambda_0}$ possesses a
conjugate point to $p_0$, which, for definiteness, we assume occurs at
a positive value of $\lambda$, i.e., to the future of $p_0$.  Then
there exists an open neighborhood $O \subset {\cal S}$ of $\Lambda_0$
such that for all $\Lambda = (p, k^a) \in O$, the null geodesic
$\gamma_\Lambda$ will possess a conjugate point to the future of
$p$. Furthermore, if we define the map $h: O \rightarrow M$ by
$\Lambda = (p, k^a) \mapsto h(p)$, where $h(p)$ is the first conjugate
point to the future of $p$ along $\gamma_\Lambda$, then h is
continuous at $\Lambda_0$.
\end{lem}
{\it Proof of Lemma 1}: Let $q_0$ denote the first conjugate point to
$p_0$ along $\gamma_{\Lambda_0}$ in the future direction, i.e., the conjugate
point to $p_0$ lying at the smallest positive value of affine parameter. Let
$\lambda_0 > 0$ denote the affine parameter value of $q_0$. Since the
exponential map is defined on an open subset of the tangent bundle, it
follows that there is an open neighborhood $\tilde{O} \subset {\cal
S}$ of $\Lambda_0$ and an $\epsilon > 0$ such that all null geodesics
determined by initial data in $\tilde{O}$ extend at least to affine
parameter $\lambda_0 + \epsilon$. We wish to show that given any open
neighborhood $U \subset M$ of $q_0$, there exists an open neighborhood
$O \subset \tilde{O}$ of $\Lambda_0$ such that for all $\Lambda \in
O$, the function $G$ on $\gamma_\Lambda$ defined by eq.(\ref{Gdef})
above will vanish for the first time at a point lying in $U$. By doing
so, we will establish the existence of a conjugate point to $p$ on
$\gamma_\Lambda$ and prove the continuity of the map $h$.

Define the map $H: \tilde{O} \times [0, \lambda_0 + \epsilon] \rightarrow M$ by
$(\Lambda, \lambda) \mapsto \gamma_{\Lambda}(\lambda)$. Then $H$ is
continuous. Therefore, given an open neighborhood $U \subset M$ of
$q_0$, there exists an open neighborhood $O' \subset \tilde{O}$ of
$\Lambda_0$ and a $\delta > 0$ (with $\delta < \epsilon$) such that
$H(\Lambda, \lambda) \in U$ whenever $\Lambda \in O'$ and
$|\lambda-\lambda_0|< \delta$.  Therefore, the lemma will be proven if
we can find an open neighborhood $O \subset O'$ of
$\Lambda_0$, such that for all $\Lambda \in O$, the first positive
value of $\lambda$ at which the function $G(\Lambda, \lambda)$
vanishes lies within $\delta$ of $\lambda_0$.

By hypothesis, $G(\Lambda_0, \lambda_0) = 0$ and $G(\Lambda_0,
\lambda) > 0$ for all $0 < \lambda < \lambda_0$. It follows
immediately from eq.(\ref{ray}) and the null energy condition that
$G''(\Lambda_0, \lambda) \leq 0$ for all $0 < \lambda < \lambda_0$. By
the mean value theorem applied to $G$, there must exist $\lambda_1 \in
(0, \lambda_0)$ such that $G'(\Lambda_0, \lambda_1) = - C$ for some $C >
0$. Since $G'' \leq 0$, it follows that $G'(\Lambda_0, \lambda) \leq - C$ for all
$\lambda \in [\lambda_1, \lambda_0)$. By choosing $\lambda_1$ to be
sufficiently near $\lambda_0$, we may assume without loss of
generality that $\lambda_0 - \lambda_1 < \delta$ and
\begin{equation}
\frac{G(\Lambda_0, \lambda_1)}{|G^\prime (\Lambda_0, \lambda_1)|}< \delta
\label{delta}
\end{equation}
since $G(\Lambda_0, \lambda_1) \rightarrow 0$ as $\lambda_1
\rightarrow \lambda_0$ but $|G^\prime (\Lambda_0, \lambda_1)|$ remains
bounded below by $C$.

From the continuous dependence of $G$ and its derivatives on
$\Lambda$, it follows that there exists an open neighborhood $O
\subset O'$ of $\Lambda_0$, such that for all $\Lambda \in
O$, we have (i) $G(\Lambda, \lambda) > 0$ for all $0 < \lambda <
\lambda_1$, (ii) $G'(\Lambda, \lambda_1) < 0$, and (iii) $G(\Lambda,
\lambda_1) / |G'(\Lambda, \lambda_1)| < \delta$. Since $G'' < 0$ by
eq.(\ref{ray}), it follows that for all $\Lambda \in O$, $G(\Lambda,
\lambda)$ must achieve its first zero between $\lambda_1$ and
$\lambda_1 + \delta$. Since $\lambda_0 - \delta < \lambda_1 <
\lambda_0$, this implies that $G(\Lambda, \lambda)$ must achieve its
first zero within $\delta$ of $\lambda_0$, as we desired to show.
$\Box$

\bigskip
\noindent
{\it Proof of Theorem 1}: Since $M$ is assumed to be paracompact, we
may introduce a Riemannian metric $q_{ab}$ on $M$ (see, e.g., Appendix
2 of \cite{kn}). By multiplying $q_{ab}$ by a conformal factor if
necessary, we may assume without loss of generality that $q_{ab}$ is
complete (see Thm. 17 of \cite{h}). Choose a point $x \in M$ and let
$r:M \rightarrow {\mathrm{I\!R}}$ denote the geodesic distance from $x$ in the
metric $q_{ab}$. Then $r$ is a continuous function on $M$ and for all $R >
0$, the set $B_R = \{p \in M| r(p) \leq R \}$ is compact (see Thm. 15 of
\cite{h}).

Since $M$ is null geodesically complete and satisfies the null energy
condition and null generic condition, it follows that every null
geodesic in $M$ contains a pair of conjugate points \cite{he}. As
discussed above, associated with each $\Lambda \in {\cal S}$ is a null
geodesic $\gamma_\Lambda$ determined by the initial conditions $\Lambda$.
Define $f:{\cal S} \rightarrow {\mathrm{I\!R}}$ by
\begin{eqnarray}
f(\Lambda)&=&\{ \inf R | \mbox{$B_R$ contains a connected segment of
$\gamma_\Lambda$ that} \nonumber \\
&& \mbox{includes the initial point determined by
$\Lambda$ together} \nonumber \\ 
&& \mbox{with a pair of conjugate points of
$\gamma_\Lambda$} \}
\label{f}
\end{eqnarray}

We claim that $f$ is an upper-semicontinuous function. To prove this,
we must show that given any $\Lambda_1 \in {\cal S}$ and given any
$\epsilon > 0$, there exists an open neighborhood, $O_1$, of
$\Lambda_1$ such that for all $\Lambda \in O_1$ we have $f(\Lambda)
\leq f(\Lambda_1) + \epsilon$. Let $p,q$ be conjugate points along
$\gamma_{\Lambda_1}$ such that a connected segment of $\gamma_\Lambda$
containing $p$, $q$, together with the initial point lies in $B_R$
with $R = f(\Lambda_1) + \epsilon/3$. Without loss of generality, we
may assume that $q$ is the first conjugate point to $p$ encountered
along $\gamma_{\Lambda_1}$ starting at $p$.  Let $\Lambda_0 = (p,
k^a)$ denote the initial data for $\gamma_{\Lambda_1}$ viewed as a
geodesic starting at $p$ (with its tangent re-scaled, if necessary, so
as to meet the normalization condition of (\ref{S})). With
$\gamma_{\Lambda_1}$ viewed in this manner, let $\lambda_0$ denote the
affine parameter of $q$. By the proof of Lemma 1 above, given any
$\delta > 0$, we can find an open neighborhood, $O_0 \subset {\cal
S}$, of $\Lambda_0$ such that the null geodesic starting at any
$(s,l^a) \in O_0$, will have a conjugate point to $s$ within affine
parameter $\lambda_0 + \delta$ of $s$. Choose $\delta$ to be
sufficiently small that the segment of $\gamma_{\Lambda_1}$ between
$\lambda_0$ and $\lambda_0 + \delta$ lies within the ball of radius
$f(\Lambda_1) + 2\epsilon/3$. By the continuity of the exponential
map, without loss of generality we may then assume that $O_0$ is
sufficiently small that for all $(s,l^a) \in O_0$, the entire null
geodesic segment starting at $s$ and ending at affine parameter
$\lambda_0 + \delta$ will lie in a ball of radius $f(\Lambda_1) +
\epsilon$. Finally, appealing again to the continuity of the
exponential map, we may choose $O_1$ to be such that (i) for all
$\Lambda \in O_1$, the segment of $\gamma_\Lambda$ between affine
parameter $0$ and $\lambda_1$ lies in the ball of radius $f(\Lambda_1)
+ \epsilon$, where $\lambda_1$ denotes the affine parameter at which
$\gamma_{\Lambda_1} (\lambda_1) = p$, and (ii) the initial data for
the geodesic $\gamma_\Lambda$ at affine parameter $\lambda_1$
corresponds to a point in $O_0$. It then follows that for all $\Lambda
\in O_1$, a connected segment of $\gamma_\Lambda$ containing the
initial point together with a pair of conjugate points will be
contained within the ball of radius $f(\Lambda_1) +
\epsilon$. Consequently $f(\Lambda) \leq f(\Lambda_1) + \epsilon$.

Let $K \subset M$ be compact. Let ${\cal S}_K = \{(p, k^a) \in {\cal
S}|p \in K\}$. Then ${\cal S}_K$ is compact. Therefore, since $f$ is
upper-semicontinuous, it must achieve a maximum, $\alpha$, on ${\cal
S}_K$. Let $K' = B_\alpha$. Suppose that $q,p \notin K^\prime$ and $q
\in J^{+}(p)-I^{+}(p)$. Let $\gamma$ be a causal curve with past
endpoint $p$ and future endpoint $q$. Then $\gamma$ must be a null
geodesic that does not contain any pair of conjugate points lying
between $p$ and $q$ (see, e.g., \cite{he}, \cite{w}). However, if $\gamma$
intersects $K$, then by the above argument, it necessarily contains a
pair of conjugate points lying in $K'$ and, hence, lying between $p$
and $q$. Consequently, $\gamma$ cannot intersect $K$
$\Box$.

\bigskip

As already indicated in the Introduction, Theorem 1 contains some
suggestion of a general ``time delay'' phenomena in general
relativity, since if it were possible to produce a ``time advance'' in
a compact region $K$, then one might expect a ``fastest null
geodesic'', $\gamma$, to take advantage of this by entering
$K$. However, since $K'$ could be far larger than $K$, it is difficult
to make a strong argument for this kind of interpretation of the
theorem.

It should be noted that in the case of asymptotically flat spacetimes,
Theorem 1 expresses a key aspect of the argument found in the
Penrose-Sorkin-Woolgar \cite{psw} positive mass theorem. In an
asymptotically flat spacetime, consider a sequence of points $p_n,
q_n$ with $q_n \in J^{+}(p_n)-I^{+}(p_n)$ such that both $\{p_n\}$ and
$\{q_n\}$ approach infinity as $n \rightarrow \infty$. Then Theorem 1
implies that given any compact set $K$ of the spacetime, for
sufficiently large $n$ the corresponding sequence of causal curves
$\{\gamma_n\}$ cannot enter $K$, i.e., for sufficiently large $n$, the
entire curve $\gamma_n$ must lie arbitrarily near infinity. The
Penrose-Sorkin-Woolgar theorem is obtained by showing that this
behavior is incompatible with a negative value of the mass of the
spacetime.

Theorem 1 also contains some implications not directly related to
``time delay''. The following Corollary\footnote{In accordance with
footnote \ref{gg}, the conclusions of Corollary 1 continue to hold if
the hypotheses that $(M, g_{ab})$ is null geodesically complete and
satisfies the null energy condition and the null generic condition is
replaced by the hypothesis that $(M, g_{ab})$ fails to contain a null
line.} establishes the absence of ``particle horizons'' in a class of
cosmological models.

\newtheorem{cor}{Corollary}
\begin{cor}
Let $(M, g_{ab})$ have the properties stated in the theorem, i.e.,
suppose that $(M, g_{ab})$ is null geodesically complete and satisfies
the null energy condition and the null generic condition. Suppose, in
addition, that $(M, g_{ab})$ is globally hyperbolic, with a compact
Cauchy surface $\Sigma$. Then there exist Cauchy surfaces $\Sigma_1$
and $\Sigma_2$ (with $\Sigma_2\subset I^+(\Sigma_1$)) such that if
$q \in I^+(\Sigma_2)$, then $\Sigma_1 \subset I^-(q)$.
\end{cor}
{\it Proof}: Since $(M, g_{ab})$ is globally hyperbolic, there exists
a continuous ``global time function'', $T:M \rightarrow
{\mathrm{I\!R}}$, such that each surface of constant $T$ is a Cauchy
surface (see, e.g., \cite{w}). Let $K = \Sigma$, and let $K'$ be as in
Theorem 1. Let $T_1$ and $T_2$ denote, respectively, the minmum and
maximum values of $T$ on $K'$. Let $\Sigma_1$ be any Cauchy surface
with $T < T_1$ and let $\Sigma_2$ denote the Cauchy surface $T =
T_2$. Let $q \in I^+(\Sigma_2)$, $p \in \Sigma_1$ and suppose that $p
\in \dot{I}^-(q)$. Since $(M, g_{ab})$ is globally hyperbolic,
$J^-(q)$ is closed (see, e.g., \cite{w}), so $p \in J^-(q) - I^-(q)$,
and thus there is a causal curve connecting $p$ to $q$. It follows from
Theorem 1 that this causal curve does not intersect $\Sigma$. However,
this contradicts the fact that $\Sigma$ is a Cauchy
surface. Consequently, there cannot exist a $p \in \Sigma_1$ and such
that $p \in \dot{I}^-(q)$, i.e., $\dot{I}^-(q) \cap \Sigma_1 =
\emptyset$. However, $I^-(q)$ is open, and since $\dot{I}^-(q) \cap
\Sigma_1 = \emptyset$, the complement of $I^-(q)$ in $\Sigma_1$ also
is open. Since we have $I^-(q) \cap \Sigma_1 \neq \emptyset$, and
$\Sigma_1$ is connected (since $M$ is assumed to be connected), this
implies that $\Sigma_1 \subset I^-(q)$, as we desired to show. $\Box$

\bigskip

It should be noted that de Sitter spacetime satisfies all the
hypotheses of our Corollary except the null generic condition.  It is
not difficult to verify that for any event $q$ in de Sitter spacetime,
$I^-(q)$ does not contain any Cauchy surface, i.e., de Sitter
spacetime also fails to satisfy the conclusion of our
Corollary---although it ``just barely'' fails in the sense that the
past of an observer arbitrarily far in the future will come
arbitrarily close to containing a Cauchy surface. However, our
Corollary shows that if we slightly perturb de Sitter spacetime so
that the null generic condition is satisfied, then at sufficiently
late times, any observer will be able to ``view the entire universe''.

\section{Time Delay in Spacetimes with a Timelike Conformal Boundary}

In this section, we shall prove a general theorem that has a direct
interpretation as showing that there is a ``time delay'' in
asymptotically anti-de Sitter spacetimes relative to anti-de Sitter
spacetime itself.  A similar result in the particular context of
asymptotically anti-de Sitter spacetimes has been previously obtained
by Woolgar \cite{wo}, who extended the arguments of \cite{psw} to the
asymptotically anti-de Sitter case.  

The general context of our theorem is one in which the physical
spacetime of interest, $(M,g_{ab})$, can be conformally embedded in a
spacetime $(\widetilde M, \tilde g_{ab})$, in such a way that the
boundary, $\dot{M}$, of $M$ in $\widetilde{M}$ is a timelike
hypersurface. We write $\bar{M} = M \cup \dot{M}$. Unless otherwise
stated, in this section all futures and pasts are understood as being
taken with respect to $\bar{M}$, so, for example,
\begin{eqnarray}
J^+(p) & \equiv &\{q\in\bar M|\mbox{there exists a causal curve in $\bar M$}
\nonumber \\
&& \mbox{connecting $p$ to $q \}$}
\label{J}
\end{eqnarray}
For $p \in \dot{M}$, we also shall be interested in considering the
events in $\dot{M}$ that can be reached by causal curves starting at
$p$ that lie entirely in $M$ except for their endpoints. For $p \in
\dot{M}$, we define
\begin{eqnarray}
A(p)&=&\{ r \in \dot{M}| \mbox{there exists a future directed causal curve $\lambda$
starting} \nonumber \\
&& \mbox{from $p$ and ending at $r$ satisfying 
$\lambda - p\cup r \subset M\}$}
\label{Ap}
\end{eqnarray}
We denote the boundary of $A(p)$ in $\dot M$ by $\dot A(p)$.

\begin{thm} 
Suppose $(M,g_{ab})$ can be conformally embedded in a spacetime
$(\widetilde M, \tilde g_{ab})$, so that in $M$ we have
$\tilde{g}_{ab}=\Omega^2 g_{ab}$ and on $\dot{M}$ we have $\Omega =
0$, where $\Omega$ is smooth on $\widetilde{M}$. Suppose $(M,g_{ab})$
satisfies the following conditions: (1) $(M,g_{ab})$ satisfies the
null energy condition\footnote{The null energy condition could be
replaced here by the weaker, averaged condition appearing in a theorem
of Borde \cite{b}.} and the null generic condition. (2) $\bar M$ is
strongly causal. (3) For any $p, q\in \bar M$, $J^+(p) \cap
J^-(q)$ is compact. (4) $\dot M$ is a timelike hypersurface in
$\widetilde{M}$. Let $p\in \dot M$. Then, for any $q\in \dot{A}(p)$, we
have $q\in J^+(p)-I^+(p)$. Furthermore, any causal curve in $\bar{M}$
connecting $p$ to $q$ must lie entirely in $\dot M$ and, hence, must
be a null geodesic in the spacetime $(\dot M, \tilde{g}_{ab})$.
\end{thm}
{\it Proof}: First, we claim that $A(p)$ is open in $\dot M$. Let
$r\in A(p)$. Then there exists a causal curve $\lambda$ connecting $p$
to $r$ which---apart from its endpoints---lies in $M$. If $\lambda$
were not a null geodesic, we could deform it to a timelike curve in
$M$ with the endpoints $p$ and $r$ fixed. However, if $\lambda$ is a
null geodesic, then since $\Omega = 0$ at its past and future
endpoints on $\dot{M}$, $\lambda \cap M$ is a complete null geodesic
in $(M, g_{ab})$ (see eq.(D.6) of \cite{w}). Since $(M, g_{ab})$
satisfies the null energy condition and null generic condition,
$\lambda$ must contain a pair of conjugate points. Therefore, in this
case, $\lambda$ also can be deformed to be a timelike curve in $M$
with the endpoints $p$ and $r$ fixed.  Thus $r\in I^+(p)$ and it
follows that we can find a neighborhood of $r$ in $\dot M$ which is in
$A(p)$, as we desired to show.

Now let $q \in \dot{A}(p)$. Clearly, $q \notin I^+(p)$, since
otherwise an open neighborhood of $q$ would lie in $A(p)$. To show
that $q \in J^+(p)$, let $\{q_n\}$ be a sequence of points in $ A(p)$
which converges to $q$. For each $q_n$, let $\lambda_n$ be a causal
curve connecting $p$ to $q_n$ which lies in $M$ apart from its
endpoints.  Since $q$ is a limit point of $\{\lambda_n\}$, we may
apply the ``limit curve'' lemma (see, e.g., \cite{he}, \cite{w}) to
$\widetilde{M} - p$. It follows that in $\widetilde{M}$, there exists
a past directed causal limit curve $\lambda$ through $q$ which either
is past inextendible or has a past endpoint at $p$. Let $q' \in I^+(q)
\cap \dot{M}$.  (Such a $q'$ exists because $\dot{M}$ is timelike.)
Then each $\lambda_n$ lies in $J^+(p) \cap J^-(q')$, which is compact,
and hence is closed as a subset of $\widetilde{M}$. Consequently,
since $\lambda$ is a limit curve of the sequence $\{ \lambda_n \}$, we
have $\lambda \subset J^+(p) \cap J^-(q')$. However, since $J^+(p)
\cap J^-(q')$ is compact and $\bar{M}$ is strongly causal, $\lambda$
cannot be past inextendible (see, e.g., lemma 8.2.1 of
\cite{w}). Thus, $\lambda$ must have a past endpoint at
$p$. Consequently, we have $q \in J^+(p)$, as we desired to show.

Finally, for $q \in \dot{A}(p)$, let $\tilde{\lambda}$ be any causal
curve in $\bar{M}$ connecting $p$ with $q$. Suppose there exists a
point $r \in \tilde{\lambda} \cap M$. Then there must exist an open
segment of $\tilde{\lambda}$ contained in $M$ with endpoints on
$\dot{M}$. By the same arguments as in the first paragraph of this
proof, we can deform this segment of $\tilde{\lambda}$ to a timelike
curve in $M$, keeping its endpoints fixed. We may then further deform
$\tilde{\lambda}$ to a timelike curve in $M$ connecting $p$ with $q$.
However, this contradicts the fact that $q \in \dot{A}(p)$. Thus, we
have $\tilde{\lambda} \cap M = \emptyset$, so $\tilde{\lambda}$ must
lie entirely in $\dot{M}$. Since $(\dot{M}, \tilde{g}_{ab})$ is a
spacetime in its own right, and since $q \in J^+(p) - I^+(p)$ [with
$J^+(p)$ and $I^+(p)$ now defined with respect to the spacetime
$(\dot{M}, \tilde{g}_{ab})$] it follows that $\tilde{\lambda}$ must
be a null geodesic with respect to $\tilde{g}_{ab}$.  $\Box$

\bigskip

It should be noted that anti-de Sitter spacetime satisfies all of the
hypotheses of Theorem 2 except for the null generic condition. It also
should be noted that anti-de Sitter spacetime fails to satisfy part of
the conclusion of Theorem 2: $n$-dimensional anti-de Sitter spacetime
is conformal to the region $0\leq r<\frac{1}{2}\pi$ of the Einstein
static universe
\begin{equation}
ds^2=-dt^2+dr^2+\sin^2 r \, d\Omega^2
\end{equation}
Thus, $\dot{M}$ is comprised by the $(n-2)$-sphere $r =
\frac{1}{2}\pi$. For $q \in \dot{A}(p)$ with $p$ and $q$ at antipodal
points of this sphere, then in addition to null geodesics in $\dot{M}$
that connect $p$ with $q$, there also exist null geodesics in $M$ that
connect $p$ with $q$ \cite{hi}, in contradiction with the last
sentence of Theorem 2. However, Theorem 2 implies that if we perturb
anti-de Sitter spacetime in such a way that the geometry of $\dot{M}$
is not changed and in such a way that all of the hypotheses of Theorem
2 hold (including the null generic condition), then $p$ and $q$ no
longer can be joined by a causal curve that enters $M$. Since the
boundary geometry has not changed, we have a well defined, fixed
``reference frame'' with respect to which we can compare the causal
properties of asymptotically anti-de Sitter spacetimes with that of
anti-de Sitter spacetime itself. Thus, as already indicated in the
Introduction, Theorem 2 may be interpreted as implying that the
propagation of light through an asymptotically anti-de Sitter
spacetime will always be ``delayed'' relative to propagation through
anti-de Sitter spacetime itself.

\bigskip

\noindent
{\bf Acknowledgements}

This research was supported in part by NSF grant PHY 95-14726 to the
University of Chicago.


\begin{thebibliography}{99} 

\bibitem{r} R.D. Reasenberg et al, Astrophys. J. Lett. {\bf 234}, 219
(1979).

\bibitem{w} R.M. Wald, {\it General Relativity}, University of Chicago
Press (Chicago, 1984).

\bibitem{vbl} M. Visser, B. Bassett, and S. Liberati, Nucl. Phys. {\bf
B88} (Proc. Supl.), 267 (2000), gr-qc/9810026; see also M. Visser,
B. Bassett, and S. Liberati, in {\it General Relativity and
Relativistic Astrophysics, Proceedings of the Eighth Canadian
Conference}, ed. by C.P Burgess and R.C. Meyers, (AIP Press, Melville,
New York, 1999), gr-qc/9908023.

\bibitem{o} K.D.Olum, Phys.Rev.Lett. {\bf 81}, 3567 (1998).

\bibitem{p} R. Penrose, in {\it Essays in General Relativity}, ed. by
F.J. Tipler, Academic Press (New York, 1980).

\bibitem{gx} R.P. Geroch and B.C. Xanthopoulos, J. Math. Phys. {\bf
19}, 714 (1978).

\bibitem{g} G.J. Galloway, math.DG/9909158.

\bibitem{wo} E. Woolgar, Class. Quant. Grav. {\bf 11}, 1881 (1994).

\bibitem{he} S.W. Hawking and G.F.R. Ellis, {\it The Large Scale
Structure of Space-Time}, Cambridge University Press (Cambridge,
1973).

\bibitem{fmw} E.E. Flanagan, D. Marolf, and R.M. Wald, Phys. Rev. D
(in press); hep-th/9908070.

\bibitem{kn} S. Kobayashi and K. Nomizu, {\it Foundations of
Differential Geometry}, Volume I, Interscience Publishers (New York,
1963).

\bibitem{h} N.J. Hicks, {\it Notes on Differential Geometry}, Van
Nostrand (Princeton, 1965).

\bibitem{psw} R. Penrose, R. Sorkin, and E. Woolgar, gr-qc/9301015.

\bibitem{b} A. Borde, Class. Quant. Grav. {\bf 4}, 343 (1987).

\bibitem{hi} G. Horowitz and N. Itzhaki,  JHEP {\bf 010}, 9902 (1999).

\end{thebibliography}
\end{document}